\documentclass[nonacm]{acmart}
\makeatletter                   
\def\mdseries@tt{m}             
\makeatother                    

\AtBeginDocument{%
  }
  
\DeclareMathOperator*{\E}{\mathbb{E}}
\usepackage{xcolor}
\usepackage{multirow}
\setcopyright{none}
\begin{document}

\title{Zero-Shot Style Transfer for Gesture Animation driven by Text and Speech \\ using Adversarial Disentanglement of Multimodal Style Encoding }

\author{Mireille Fares}
\affiliation{%
  \institution{ISIR, STMS, Sorbonne University}
  \city{Paris}
  \country{France}}
\email{fares@isir.upmc.fr}

\author{Michele Grimaldi}
\affiliation{%
  \institution{ISIR, Sorbonne University}
  \city{Paris}
  \country{France}}
\email{grimaldi@isir.upmc.fr}

\author{Catherine Pelachaud}
\affiliation{%
  \institution{CNRS, ISIR, Sorbonne University}
  \city{Paris}
  \country{France}}
\email{catherine.pelachaud@sorbonne-universite.fr}

\author{Nicolas Obin}
\affiliation{%
  \institution{STMS, Sorbonne University}
  \city{Paris}
  \country{France}}
\email{nobin@ircam.fr}

\begin{abstract}
Modeling virtual agents with behavior style is one factor for personalizing human-agent interaction.  In this paper, we propose an efficient yet effective machine learning approach to synthesize gestures driven by prosodic features and text in the style of different speakers including those unseen during training. Our model performs zero-shot multimodal style transfer driven by multimodal data from the PATS database containing videos of various speakers. We view style as being pervasive while speaking; it colors the communicative behaviors expressivity while speech content is carried by multimodal signals and text. This disentanglement scheme of content and style allows us to directly infer the style embedding even of speaker whose data are not part of the training phase, without requiring any further training or fine-tuning. The first goal of our model is to generate the gestures of a source speaker based on the \emph{content} of two input modalities – Mel spectrogram and text semantics. The second goal is to condition the source speaker’s predicted gestures on the multimodal behavior \emph{style} embedding of a target speaker. The third goal is to allow zero-shot style transfer of speakers unseen during training without re-training the model. Our system consists of two main components: (1) a \emph{speaker style encoder network} that learns to generate a fixed-dimensional speaker embedding \emph{style} from a target speaker multimodal data (mel-spectrogram, pose, and text); and (2) a \emph{sequence-to-sequence synthesis network} that synthesizes gestures based on the \emph{content} of the input modalities - text and mel-spectrogram - of a source speaker, and conditioned on the speaker style embedding.  We evaluate that our model is able to synthesize gestures of a source speaker given the two input modalities, and transfer the knowledge of target speaker style variability learned by the speaker style encoder to the gesture generation task in a zero-shot setup, indicating that the model has learned a high quality speaker representation. For our evaluation we convert the 2D generated gestures to 3D poses, and produce 3D animations of the generated gestures. We conduct objective and subjective evaluations to validate our approach and compare it with baselines.

\end{abstract}


\maketitle
\thispagestyle{empty}
\pagestyle{empty}

\section{Introduction}
Human behavior style is a socially meaningful clustering of features found within and across multiple modalities, specifically in linguistic \cite{campbell2006elements}, spoken behavior such as the speaking style conveyed by speech prosody \cite{moon2022mist, obin2011melos}, and nonverbal behavior such as hand gestures and body posture \cite{obermeier2015speaker, wagner2014gesture}. Style involves the ways in which people talk differently in different situations. A same person may have different speaking styles depending on the situation (e.g. at home, at the office or with friends). These situations  can carry different social meanings \cite{bell1984language}. Different persons may also have different behavior styles while communicating in  similar contexts. 
Style is syntagmatic. It unfolds over time in the course of an interaction and during one’s life course \cite{campbell2006elements}. It does not emerge unaltered from the speaker. It is continuously attuned as it is accomplished and co-produced with the audience \cite{mendoza1999style}. It can be very self-conscious and at the same time can be extremely routinized to the extent that it resists attempts of being altered \cite{mendoza1999style}. For instance, style-shifting has been observed in the speech of Oprah Winfrey \cite{hay1999oprah}, a popular host of a U.S. talk show. Internal linguistic factors such as lexical frequency, and external sociolinguistic factors influence the phonetic of various variables in her speech \cite{hay1999oprah}. Another study \cite{putra2019analysis} shows that Ellen Degeneres, another popular host of a US talk show, employs different speech styles in her TV show such as formal, consultative, casual and intimate styles.  Style is specifically related to the diversity of gestures and expressivity of each specific speaker \cite{pelachaud2009studies, bergmann2009increasing}. All of the aforementioned points constitute a technical challenge when trying to model behavior style in virtual agents. The behavior generation model should not simply learn an overall style from multiple speakers, but should remember each speaker's specific style - idiosyncrasy - generated in a specific lexical content context and behavior expressivity. The model should be able to capture the style that are common throughout speakers, the ones that are unique to a speaker's prototypical gestures produced consciously and unconsciously, as well as the different style-shifting that may occur during speech. 

Verbal and non-verbal behavior play a crucial role in sending and perceiving new information \cite{36} in human-human interaction. Generative models that aim to predict Embodied Conversational Agents (ECA) gestures must consider the importance of producing meaningful and naturalistic gestures that are aligned with speech \cite{cassell00nudge}. Non-verbal behavior must be generated and synchronized in conjunction with verbal and prosodic behavior to define their shape and time of occurrence \cite{salem}. This constitutes another technical challenge, to enable a smooth and engaging interaction between humans and ECAs by making sure that ECAs produce semantically-aware, natural, expressive and coherent gestures aligned with speech and its content. 
 
In the present paper, we propose a novel approach to model behavior style in virtual agents and to tackle the different style modeling 
challenges. Our approach aims at (1) synthesizing natural and expressive upper body gestures of a source speaker, by encoding the \emph{content} of two input modalities – text semantics and Mel spectrogram, (2) conditioning the source speaker’s predicted gesture on the multimodal \emph{style} representation of a target speaker, and therefore rendering the model able to perform style transfer across speakers, and finally (3) allowing zero-shot style transfer of newly coming speakers that were not seen by the model during training. Our model consists of two main components: first (1) a speaker style encoder network which goal is to model a specific target speaker style extracted from three input modalities – Mel spectrogram, upper-body gestures, and text semantics; and second (2) a sequence-to-sequence synthesis network that generates a sequence of upper-body gestures based on the content of two input modalities – Mel spectrogram and text semantics – of a source speaker, and conditioned on the target speaker style embedding. We trained our model on the database PATS, which was proposed in \cite{ahuja2020style} and designed to study gesture generation and style transfer. It includes 3 main features that we are considering in our approach: text semantics represented by BERT embeddings, Mel spectrogram and 2D upper body poses.
\subsection{Contributions}
Our contributions can be listed as follows:
\begin{enumerate}
    \item We propose the first approach for zero-shot multimodal style transfer approach for 2D pose synthesis. At inference, an embedding style vector can be directly inferred from multimodal data (text, speech and and pose) of any speaker, by simple projection into the embedding style space (similar to the one used in \cite{jia2018transfer}). The style transfer performed by our model allows the transfer of style from any unseen speakers, without further training or fine-tuning of our trained model. Thus it is not limited to the styles of the speakers of a given database. It also allows \emph{"style preservation"} by generating gestures for multiple speakers while remembering what is unique for each speaker.    
    \item To design our approach, we make the following assumptions for the separation of style and content information: \emph{style} is possibly encoded across all modalities (text, speech, pose) and varies little or not over time; \emph{content} is encoded only by text and speech modalities and varies over time.
    \item To implement theses assumptions, we propose an architecture for encoding and disentangling \emph{content} and \emph{style} information from multiple modalities. On one side, a content encoder is used to encode a content matrix from text and speech signal; on the other hand, a style encoder is used to encode a style vector from all text, speech, and signal modalities. A fader loss is introduced to effectively disentangle content and style encodings \cite{lample2017fader}.
The encoding of the style takes into account  3 modalities: body poses, text semantics, and speech - Mel spectrograms. These modalities are important 
for gesture generation \cite{kucherenko2019analyzing, Ginosar_2019_CVPR} and are linked to style. 
    \item  Finally, we evaluate the 2D generated gestures by converting them to 3D poses, and simulating 3D animations of the generated gestures. The 3D poses generation is done from incomplete upper body 2D pose joints, using MocapNET, and are simulated on a 3D virtual agent. 3D poses estimation has never been done using 2D poses with such a large number of missing joints in the context of virtual agents animation. 
\end{enumerate}
 The paper is organized as follows. The next section discusses the related work. We then describe the proposed architecture. Afterwards we describe the training regime. Then we present the objective and subjective evaluations. We finally discuss our results.
 
\section{Related Work}
Since few years, a large number of gesture generative models have been proposed,  principally based on sequential generative parametric models such as Hidden Markov Models HMM and gradually moving towards deep neural networks enabling spectacular advances over the last few years. Hidden Markov Models were previously used to predict head motion driven by prosody \cite{sargin2008analysis}, and body motion \cite{levine2009real, marsella2013towards}. 
Chiu \& Marsella \cite{chiu2014gesture} proposed an approach for predicting gesture labels from speech using conditional random fields (CRFs) and generating gesture motion based on these labels, using Gaussian process latent variable models (GPLVMs). These works focus on the gesture generation task driven by either one modality namely speech, or by the two modalities - speech and text. Their work focuses on producing naturalistic and coherent gestures that are aligned with speech and text, enabling a smoother interaction with ECAs, and leveraging the vocal and visual prosody. The non-verbal behavior is therefore generated in conjunction with the verbal behavior. Most of these works use a TTS for producing the voice, which, then, serves as input for computing the animation of the virtual agent.  LSTM networks driven by speech were recently used to predict sequences of gestures \cite{hasegawa2018evaluation} and body motions \cite{shlizerman2018audio, ahuja2019react}. LSTMs were additionally employed for synthesizing sequences of facial gestures driven by text and speech, namely the fundamental frequency (F0)\cite{fares2020towards, fares2021multimodalwacai}. Generative adversarial networks (GANs) were  proposed to generate realistic head motion \cite{sadoughi2018novel} and body motions \cite{ferstl2019multi}. Furthermore, transformer networks and attention mechanisms were recently used for upper-facial gesture synthesis based on multimodal data - text and speech \cite{fares2021multimodal}. Jonell et al. \cite{jonell2020let} propose a probabilistic approach based on normalizing flows for synthesizing facial gestures in dyadic settings.
Gestures driven by both acoustic and semantic information \cite{fares2021multimodal, kucherenko2020gesticulator,fares2020towards} are the closest approaches to our gesture generation task, however they cannot be used for the style transfer task.

Beyond realistic generation of human non-verbal behavior, style modelling and control in gesture is receiving more attention in order to propose more expressive behaviors that could possibly adapted to a specific audience \cite{neff2008gesture, karras2017audio, cudeiro2019capture, ahuja2020style, Ginosar_2019_CVPR, alexanderson2020style, Ahuja_CVPR_lowRes}. Michael Neff et al.\cite{neff2008gesture} propose a system that produces full body gesture animation driven by text, in the style of a specific performer. Alexanderson et al. \cite{alexanderson2020style} propose a generative model for synthesizing speech-driven gesticulation, they exert directorial control over the output style such as gesture level and speed. Tero Karras et al.\cite{karras2017audio} propose a model for driving 3D facial animation from audio. Their main objective is to model the style of a single actor by using a deep neural network that outputs 3D vertex positions of meshes that correspond to a specific audio. Daniel Cudeiro et al.\cite{cudeiro2019capture} also propose a model that synthesizes 3D facial animation driven by speech signal. Ginosar et al. \cite{Ginosar_2019_CVPR} propose an approach for generating gestures given audio speech, however their approach uses models trained on single speakers. The aforementioned works have focused on generating nonverbal behaviors (facial expression, head movement, gestures in particular) aligned with speech \cite{neff2008gesture, karras2017audio, cudeiro2019capture, ahuja2020style}. They have not consider multimodal data when modeling style, as well as when synthesizing gestures. 


To our knowledge, the only attempts to model and transfer the style from multi-speakers database have been proposed by \cite{ahuja2020style} and \cite{Ahuja_CVPR_lowRes}. \cite{ahuja2020style} presented Mix-StAGE, a speech driven approach that trains a model from multiple speakers while learning a unique style embedding for each speaker. They created PATS, a dataset designed to study various styles of gestures for a large number of speakers in diverse settings. 
In their proposed neural architecture, a content and a style encoder are used to extract content and style information from speech and pose. To disentangle style from content information, they assume that style is only encoded through the pose modality, and the content is shared across speech and pose modalities. A style embedding matrix whose each vector represents the style associated to a specific speaker from the training set.
During training, they further propose a multimodal GAN strategy to generate poses either from the speech or pose modality. During inference, the pose is inferred by only using the speech modality and the desired style token.
However, their generative model is conditioned on gesture style and driven by audio. It does not include verbal information. It cannot perform zero-shot style transfer on speakers that were not seen by their model during training. In addition, the style is associated with each unique speaker, which makes the distinction unclear between each speaker's specific style - idiosyncrasy -, the style that is shared among a set of speakers of similar settings (i.e. TV show hosts, journalists, etc...), and the style that is unique to each speaker's prototype gestures that are produced consciously and unconsciously, in addition to  the different style-shifting that may occur. Moreover, the style transfer is limited to the styles of the speakers of , which prevents the transfer of style from an unseen speaker. Furthermore, the proposed architecture is based on the disentangling of content and PATS style information, which is based on the assumption that style is only encoded by gestures. However, both text and speech also convey style information, and the encoding of style must take into account all the modalities of human behavior. To tackle those issues, \cite{Ahuja_CVPR_lowRes} presented a few-shot style transfer strategy based on neural domain adaptation accounting for cross-modal grounding shift between source speaker and target style. This adaptation still requires 2 minutes of the style to be transferred.

To the best of our knowledge, our approach is the first to synthesize  gestures from a source speaker, which are semantically-aware, speech driven and conditioned on a multimodal representation of the style of target speakers, in a zero-shot configuration i.e., without requiring any further training or fine-tuning. 

\section{Zero-Shot Multimodal Style Transfer Model (ZS-MSTM) for Gesture Animation driven by Text and Speech}
\begin{figure}[H]
\includegraphics[width=\textwidth]{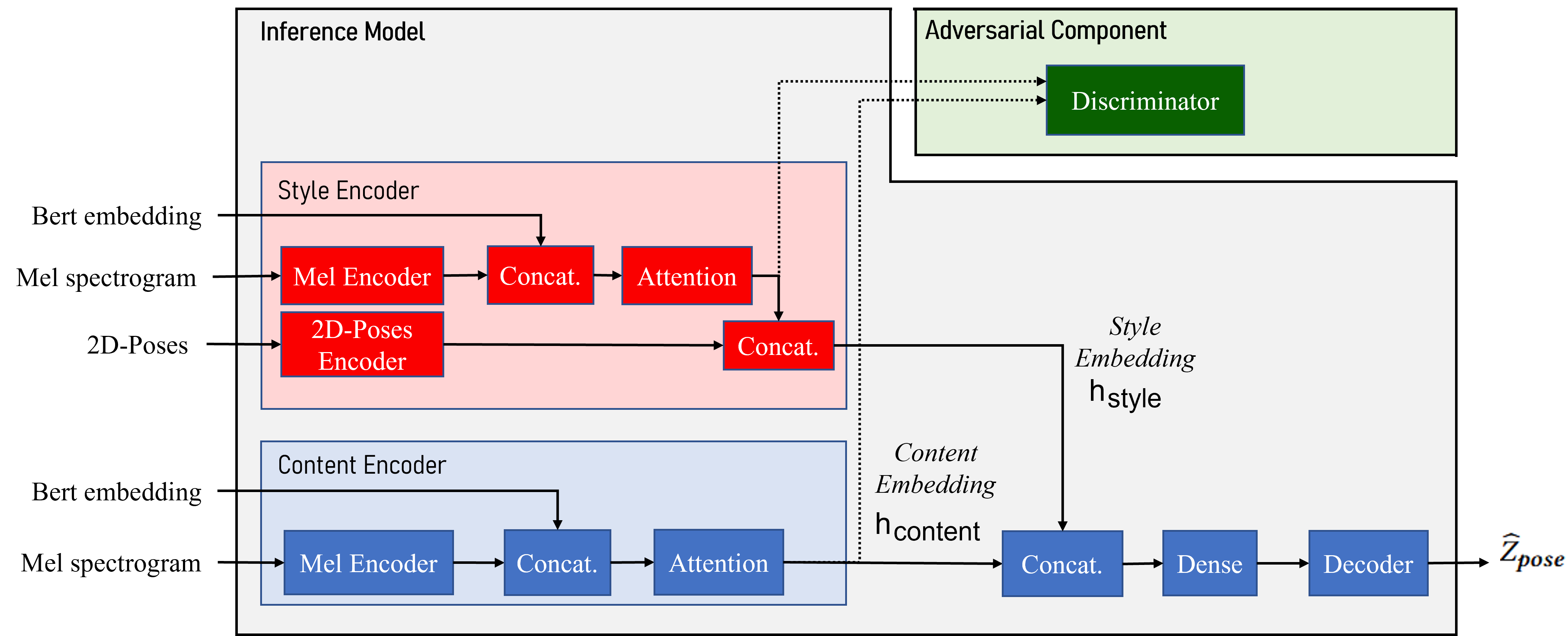}
\caption{ \textbf{ZS-MSTM} (\textbf{Z}ero-\textbf{S}hot \textbf{M}ultimodal \textbf{S}tyle \textbf{T}ransfer \textbf{M}odel) architecture. The content encoder (further referred to as $E_{content}$) is used to encode content embedding $h_{content}$ from  BERT text embeddings $X_{text}$ and speech Mel-spectrograms $Y_{speech}$ using a speech encoder $E_{content}^{speech}$. The style encoder (further referred to as $E_{style}$) is used to encode style embedding $h_{style}$ from multimodal text $X_{text}$, speech $Y_{speech}$, and pose $Z_{pose}$ using speech encoder  $E_{style}^{speech}$ and pose encoder $E_{style}^{pose}$. The generator $G$ is a transformer network that generates the sequence of poses $\widehat{Z}_{pose}$ from the sequence of content embedding $h_{content}$ and the style embedding vector $h_{style}$. The adversarial module relying on the discriminator $D$ is used to disentangle content and style embeddings $h_{content}$ and $h_{style}$. }
\label{fig:ArchitectureOverview}
\vspace{-0.3cm}
\end{figure}
We propose \textbf{ZS-MSTM} (\textbf{Z}ero-\textbf{S}hot \textbf{M}ultimodal \textbf{S}tyle \textbf{T}ransfer \textbf{M}odel), a transformer-based architecture for stylized upper-body gesture synthesis, driven by the content of a source speaker's speech - text semantics represented by BERT embeddings and audio Mel spectrogram -, and conditioned on a target speaker's multimodal style embedding. The stylized generated gestures correspond to the style of target speakers seen and unseen during training. 
As depicted in Fig.\ref{fig:ArchitectureOverview}, the system is composed of three main components:
\begin{enumerate}
    \item A \textbf{speaker style encoder} network that learns to generate a fixed-dimensional speaker embedding style from a \emph{target speaker} multimodal data: 2D poses, BERT embeddings, and Mel spectrogram, all extracted from videos in a database.
    \item A \textbf{sequence to sequence gesture synthesis} network that synthesizes gestures based on the content of two input modalities - text embeddings and Mel spectrogram - of a \emph{source speaker}, and conditioned on the \emph{target speaker} style embedding. A \emph{content encoder} is presented to encode the content of the Mel spectrogram along with BERT embeddings.
    \item An \textbf{adversarial component} in the form of a fader network \cite{lample2017fader} is used for disentangling style and content from the multimodal data. 
\end{enumerate}

At inference time, the adversarial component is discarded, and the model can generate different versions of poses when fed with different style embeddings. Gesturing styles for the same input speech can be altered by simply switching the style embeddings, or switching the multimodal input data fed as input to the Style Encoder. 

ZS-MSTM illustrated in Fig. \ref{fig:ArchitectureOverview} aims at mapping multimodal speech and text feature sequences into continuous upper-body gestures, conditioned on a speaker style embedding. The network operates on the word-level: the inputs and output of the network consist of one feature vector for each word \emph{\textbf{W}} of the input text sequence. The length of the word-level input features (text and audio) corresponds to \emph{64} timesteps (as provided by PATS). The model generates a sequence of gestures corresponding to the same word-level features given as inputs. Gestures are sequences of 2D poses represented by \emph{X} and \emph{Y} positions of the joints of the skeleton. The network has an embedding dimension \emph{d$\_$model} equal to 768. 
\subsection{Content Encoder}

The content encoder $E_{content}$ illustrated in Fig.\ref{fig:ArchitectureOverview} takes as inputs BERT embedding $X_{text}$ and audio Mel spectrograms $Y_{speech}$ corresponding to each \textbf{W}.  $X_{text}$ is represented by a vector of length 768 - BERT embedding size used in PATS. $Y_{speech}$ is encoded using \emph{Mel spectrogram Transformer (AST)} pre-trained \emph{base384} model \cite{gong2021ast}. 
\emph{AST} operates as follows: the input Mel spectrogram which has 128 frequency bins, is split into a sequence of 16x16 patches with overlap, and then is linearly projected into a sequence of 1D patch vectors, which is added with a positional embedding. We append a [\emph{CLS}] token to the resulting sequence, which is then input to a \emph{Transformer Encoder}. \emph{AST} was originally proposed for audio classification. Since we do not intend to use it for a classification task, we remove the linear layer with sigmoid activation function at the output of the \emph{Transformer Encoder}. We use the \emph{Transformer Encoder}'s output of the [\emph{CLS}] token as the Mel spectrogram representation \textbf{S}. The \emph{Transformer Encoder} has an embedding dimension equals to $d_{model}$, 12 encoding layers, and 12 attention heads.
The word-level encoded Mel spectrogram is then concatenated with the word-level BERT embedding. A self-attention mechanism is then applied on the resulting vector. The multi-head attention layer has 4 attention heads, and an embedding size $d_{att}$ equals to $d_{att}=d_{model}+768$. The output of the attention layer is the vector $h_{content}$, a content representation of the source speaker's word-level Mel spectrogram and text embedding, and it can be written as follows:
\begin{equation} \label{eqn1}
	h_{content} =  sa\left( \left[ E_{content}^{speech} (Y_{speech}),  X_{text} \right] \right)
\end{equation}
where: sa(.) denotes self-attention.
\subsection{Style Encoder}
As discussed previously, \emph{style} is a clustering of features found within and across modalities, encompassing verbal and non-verbal behavior. It is not limited to gestural information. We consider that style is  encoded in a speaker's multimodal - text, speech and pose - behavior. As illustrated in Fig.\ref{fig:ArchitectureOverview}, the style encoder $E_{style}$  takes as input, at the word-level,  Mel spectrogram $Y_{speech}$, BERT embedding $X_{text}$, and a sequence of (X, Y) joints positions that correspond to a target speaker's 2D poses $Z_{pose}$. \emph{AST} is used to encode the audio input spectrogram. 3 layers of LSTMs with a hidden-size equal to $d_{model}$ are used to encode the vector representing the 2D poses. The last hidden layer is then concatenated with the audio representation. Next, a multi-head attention mechanism is  applied on the resulting vector. This attention layer has 4 attention heads and an embedding size equals to $d_{att}$. Finally, the output vector is  concatenated with the 2D poses vector representation. The resulting vector $h_{style}$  is the output speaker style embedding that serves to condition the network with the speaker style. The final style embedding $h_{style}$ can therefore be written as follows :
\begin{equation} \label{eqn2}
		h_{style} = \left[ sa  \left( \left[X_{text}, E_{style}^{speech}  (Y_{speech}) \right] \right),  E_{style}^{pose} (Z_{pose})   \right] 
\end{equation}
where: sa(.) denotes self-attention.
\subsection{Sequence to sequence gesture synthesis}
The stylized 2D poses are generated given the sequence of content representation $h_{content}$ of the source speaker's  Mel spectrogram and text embeddings obtained at word-level, and conditioned by the style vector embedding $h_{style}$   generated from a target speaker's multimodal data. For decoding the stylized 2D-poses, the sequence of $h_{content}$ and the vector $h_{style}$ are concatenated (by repeating the $h_{style}$ vector for each word of the sequence), and passed through a dense layer of size $d_{model}$. 
We then give the resulting vector as input to a transformer decoder. The transformer decoder is composed of \emph{N = 1} decoding layer, with 2 attention heads, and an  embedding size equal to $d_{model}$. Similar to the one proposed in \cite{vaswani2017attention}, it is composed of residual connections applied around each of the sub-layers, followed by layer normalization. Moreover, the self-attention sub-layer in the decoder stack is altered  to prevent positions from attending to subsequent positions. The output predictions are offset by one position. This masking makes sure that the predictions for position index \emph{\textbf{j}} depends only on the known outputs at positions that are less than \emph{\textbf{j}}. For the last step, we  perform a permutation of the first and the second dimensions of the vector generated by the transformer decoder. The resulting vector is a sequence of 2D-poses which corresponds to: 
\begin{equation}
\widehat{Z}_{pose} = G(h_{content}, h_{style})
\end{equation}
where: G is the transformer generator conditioned on latent content embedding $h_{content}$ and style embedding $h_{style}$ . 
The generator loss of the transformer gesture synthesis can be written as,
\begin{equation} \label{eqn3}
	\mathcal{L}^{gen}_{rec}(E_{content}, E_{style}, G) = \E_{\widehat{Z}_{pose}} ||\widehat{Z}_{pose} - G(h_{content}, h_{style}) ||_2
\end{equation}
\vspace{-1cm}
\subsection{Adversarial Component}
Our approach of disentangling style from content relies on the fader network disentangling approach \cite{lample2017fader}, where a fader loss is introduced to effectively separate content and style encodings. The fundamental feature of our disentangling scheme is to constrain the latent space of $h_{content}$ to be independent of the style embeddings $h_{style}$. Concretely, it means that the distribution over $h_{content}$ of the latent representations should not contain the style information. A fader network is composed of: an encoder which encodes the input information \emph{X} into the latent code $h_{content}$, a decoder which decodes the original data from the latent, and an additional variable $h_{style}$ used to condition the decoder with the desired information (a face attribute in the original paper). The objective of the fader network is to learn a latent encoding $h_{content}$ of the input data that is independent on the conditioning variable $h_{style}$ while both variables are complementary to reconstruct the original input data from the latent variable $h_{content}$ and the conditioning variable $h_{style}$. To do so, a discriminator \emph{D} is optimized to predict the variable $h_{style}$ from the latent code $h_{content}$; on the other side the auto-encoder is optimized using an additional adversarial loss so that the classifier \emph{D} is unable to predict the variable $h_{style}$. Contrary to the original fader network in which the conditional variable is discrete within a finite binary set (0 or 1 for the presence or absence attribute), in this paper the conditional variable $h_{style}$ is continuous. We then formulate this discriminator as a regression on the conditional variable $h_{style}$: the discriminator learns to predict the style embedding $h_{style}$ from the content embedding $h_{content}$, as:
\begin{equation}
\widehat{h}_{style} = D({h}_{content})
\end{equation}
While optimizing the discriminator, the discriminator loss $\mathcal{L}^{dis}$ must be as low as possible, such as:
\begin{equation}
\mathcal{L}^{dis}(D) = \E_{\widehat{h}_{style}} ||{h}_{style} - D(h_{content}) ||_2
\end{equation}
In turn, optimizing the generator loss including the fader loss $\mathcal{L}^{gen}_{adv}$, the discriminator must not be able to predict correctly the style embedding $h_{style}$ from the content embedding $h_{content}$ conducting to a high discriminator error and thus a low fader loss. 
The adversarial loss can be written as,
\begin{equation}
\mathcal{L}^{gen}_{adv}(E_{content}, E_{style}, G) \E_{\widehat{h}_{style}} || 1 - (h_{style} - D(h_{content})) ||_2
\end{equation}
To be consistent, the style prediction error is preliminary normalized within 0 and 1 range.
\begin{figure}
\includegraphics[width=9cm]{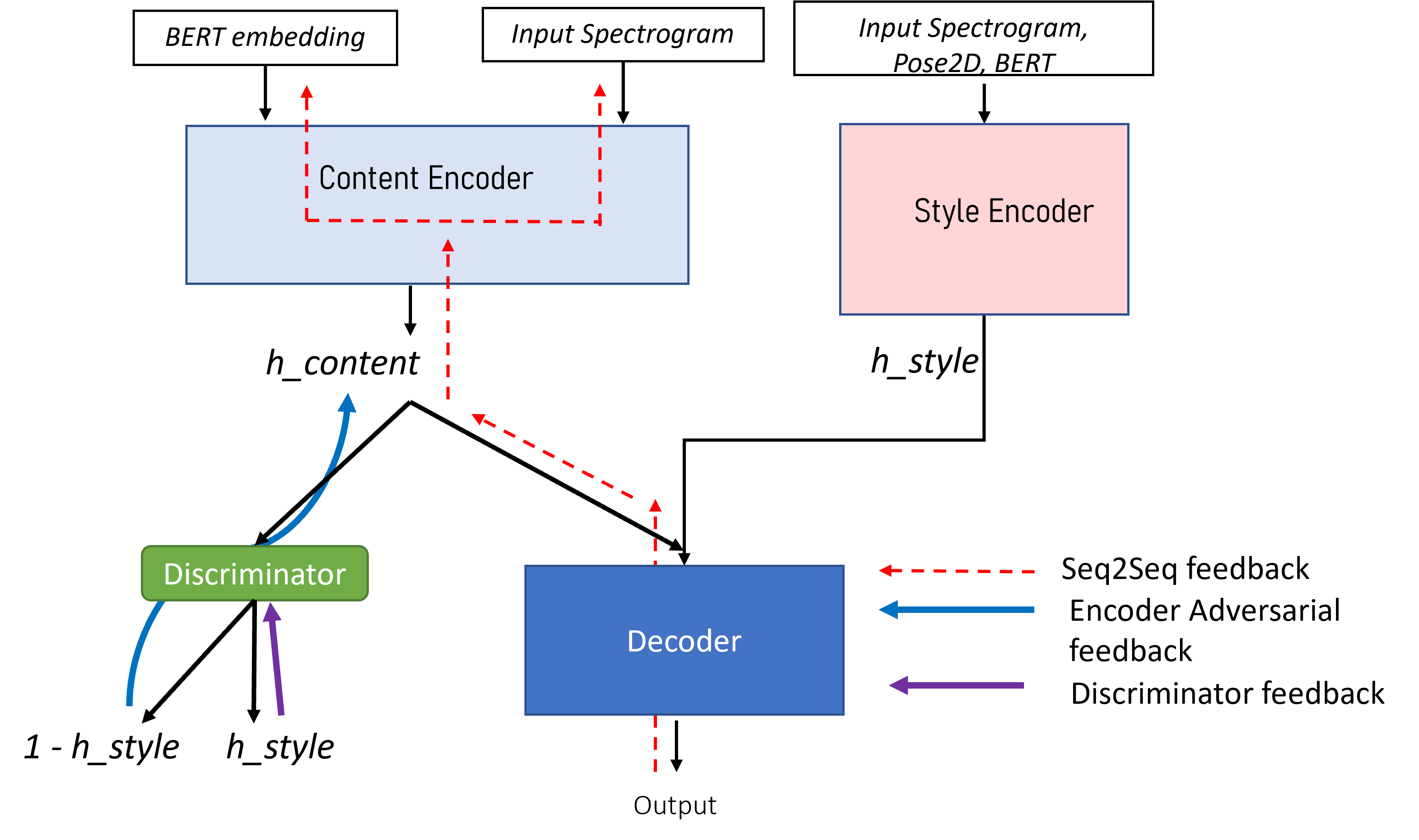}
\caption{Fader network for multimodal content and style disentangling.}
\label{fig:FaderNetwork}
\vspace{-0.2cm}
\end{figure}


\noindent Finally, the total generator loss can therefore be written as follows:
\begin{equation} \label{eqn4}
	\mathcal{L}^{gen}_{total}(E_{content}, E_{style}, G)  = \mathcal{L}^{gen}_{rec}(E_{content}, E_{style}, G) + \lambda \mathcal{L}^{gen}_{adv}(E_{content}, E_{style}, G)
\end{equation}
where $\lambda$ is the adversarial weight that starts off at 0 and is linearly incremented by 0.01 after each training step. 
The discriminator $D$ and the generator $G$ are then optimized alternatively as described in \cite{lample2017fader}.
\section{Training}
This section describes the training regime we follow for our model. 
We trained our network using the PATS dataset \cite{ahuja2020style}. PATS was created to study various styles of gestures. The dataset contains upper-body 2D pose sequences aligned with corresponding Mel spectrogram, and BERT embeddings. It offers 251 hours of data, with a mean of 10.7 seconds and a standard deviation of 13.5 seconds per interval. PATS gathers data from 25 speakers with different behavior styles from various settings (e.g., lecturers, TV shows hosts). It contains also several annotations. The spoken text has been transcribed in PATS and aligned with the speech. The 2D body poses have been extracted with OpenPose.  Each speaker is represented by their lexical diversity and the spatial extend of their arms. While in PATS arms and fingers have been extracted, we do not consider finger data in our work. That is we do not model and predict 2D finger joints. This choice arises as the analysis of finger data is very noisy and not very accurate. 

We consider two test conditions: \emph{Seen Speaker} and \emph{Unseen Speaker}. The \emph{Seen Speaker} condition aims to assess the style transfer correctness that our model can achieve when presented with speakers that were seen during training as target style. On the other hand, the \emph{Unseen Speaker} condition aims to assess the performance of our model when presented with unseen target speakers, to perform zero-shot style transfer. Seen and unseen speakers are specifically selected from PATS to cover a diversity of stylistic behavior with respect to lexical diversity and spatial extent as reported by \cite{ahuja2020style}\footnote{https://chahuja.com/pats/}. For each PATS speaker, there is a train, validation and test set already defined in the database. For testing the \emph{Seen Speaker} condition, our training set includes the train sets of 16 PATS speakers: "Shelly", "Jon", "Fallon", "Bee", "Ellen", "Oliver", "Lec\_cosmic", "Lec\_hist", "Seth", "Conan", "Angelica", "Rock", "Noah", "Ytch\_prof", "Lec\_law", and "Ytch\_dating". Six other speakers are selected for the \emph{Unseen Speaker} condition, and their test sets are also used for our experiments. These six speakers "Lec\_evol", "Almaram", "Huckabee", "Ytch\_charisma", "Minhaj", and "Chemistry" differ in their behavior style and lexical diversity.  
Each training batch contains 24 pairs of word embeddings, Mel spectrogram, and their corresponding sequence of (X, Y) joints of the skeleton (of the upper-body pose). We use Adam optimizer with $\beta_{1} = 0.95, \beta_{2} = 0.999$. For balanced learning, we use a scheduler with an initial learning rate of 0.00001, with \emph{warmup steps = 20000}. We train the network for 200 epochs. All features values are normalized so that the dataset mean and standard deviation are 0 and 0.5, respectively.
\section{3D Pose Generation and Simulation}
Previous evaluation studies of models learned from video data have used 2D stick figures for their subjective evaluation \cite{ahuja2020style}. Even when the 2D stick figure is projected on the video of a human speaker, the animation is not always readable as, in particular, it is missing information on the body pose in the Z direction (the depth axis). So we choose to convert the 2D poses into 3D poses. We visualize the behavior animation resulting from our model on a 3D virtual agent. As in \cite{ahuja2020style}, we train our model on the database PATS, and therefore the generated 2D body poses correspond to incomplete skeleton joints; missing joints include lower body joints, as well as torso joints. To visualize the resulting animations of our model, we convert the 2D poses into 3D poses and use 3D human mesh.

We develop an approach that generates 3D poses from incomplete upper body 2D pose joints using  MocapNET \cite{Qammaz2019}, an ensemble of SNN encoders that estimates the 3D human body pose based on 2D joint estimations extracted from monocular RGB images. It outputs skeletal information directly into the BVH format which can be rendered in real-time or imported without any additional processing in most popular 3D animation software.
MocapNET operates on 2D joint input, received in the popular COCO\cite{cao2017realtime} or BODY25\cite{cao2017realtime} format. In order to be used, the file containing the predictions are formatted following  the BODY25 format and the 2D joints are mapped to respect the BODY25  joints. The JSON files with 2D detections are subsequently converted to CSV files and then to 3D BVH files using the MocapNET. Finally we add zeros for the missing joints.
MocapNET is trained using a 1920x1080 "virtual camera" to emulate a GoPRO Hero 4 running at the Full-HD mode. We adapted the output of our gesture generation model to such a configuration. We also set up  the  frames resolution to correspond to the original video  stream size.
\begin{figure}
\centering
\includegraphics[width=10cm]{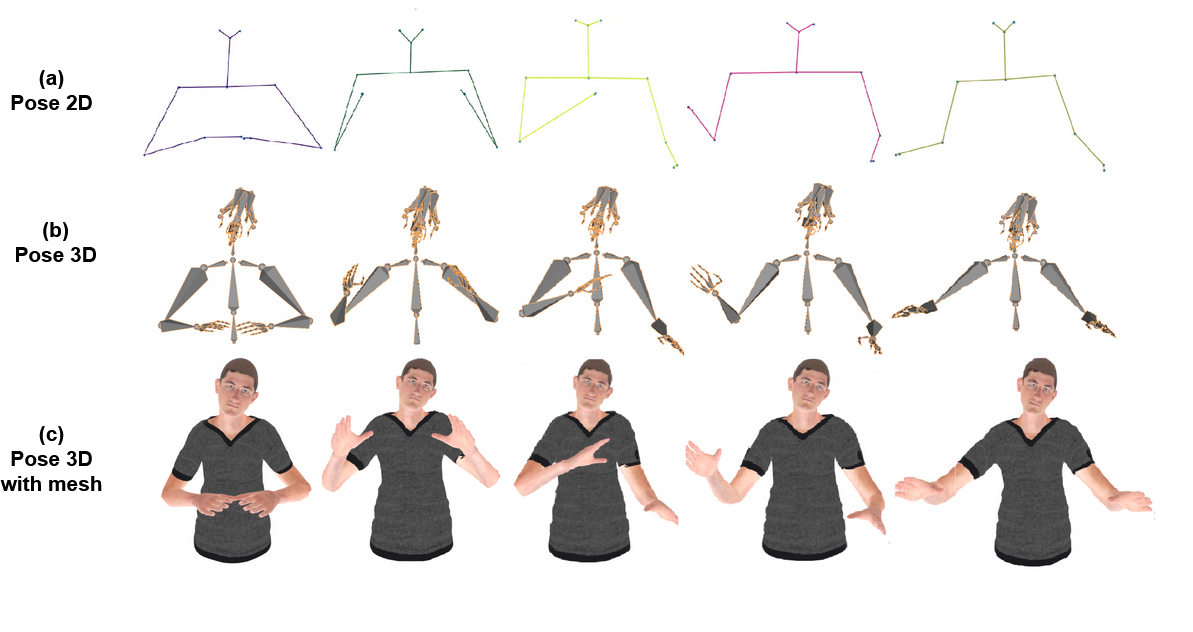}
\caption{A sequence of gestures corresponding to a sequence of 2D poses. (a) 2D poses. (b) The corresponding sequence of 3D poses computed by MocapNet and simulated with Blender. (c) Resulting animation with a 3D human mesh.}
\vspace{-0.3cm}
\end{figure}
Once the BVH file is created we use the 3D animation software Blender to simulate the animation.  Finally, we apply a 3D human mesh to the skeleton to simulate a 3D human animation. The mesh is taken from Mixamo\footnote{https://www.mixamo.com/}, an online database of characters and mocap animations used in art projects, movies and games. In order to fuse the mesh with the skeleton, we scale the mesh to fit the skeleton and we parent the skeleton and the mesh with automatic weights. 

\section{Evaluation Metrics and Studies}
To measure the performance of our work, we conducted several objective and subjective evaluation studies we present in this section. We start by introducing the metrics we use for the objective studies; we follow by explaining the protocol we follow for the perceptive studies as well as the creation of stimuli we use.

\subsection{Objective Evaluation Metrics}
In our work, we have defined style by the behavior expressivity of a speaker. To evaluate objectively our works, we define metrics to compare the behavior expressivity of a speaker from the ground truth with the predicted behavior expressivity generated by our model in different conditions.
Following works on behavior expressivity by \cite{Wallbott98,pelachaud2009studies}, we define 4 objective behavior dynamics metrics to evaluate the style transfer of different target speakers:  Acceleration, jerk and   velocity that are averaged over the values of all upper-body joints, as well as the speaker's average bounding box perimeter of his/her body movements extension.
In addition, we compute the acceleration, jerk and velocity of only the left and right wrists, to obtain information on the arms movements expressivity \cite{Wallbott98,kucherenko2019analyzing}.

\subsection{Subjective Evaluation}
We also conduct three human perceptual studies.
As a pre-evaluation of our approach, we conduct a human perceptual study (\textbf{Study 1}) to validate the 2D to 3D pose conversion by measuring the \emph{resemblance} of the 3D animations to the ground truth in terms of the expressivity of the style (gesture amplitude and dynamics), in addition to the quality of the produced 3D animations (such as naturalness and comprehensibility of movements). 
Then, to investigate human perception of the stylized upper-body gestures produced by our model, we conduct another human perceptual study (\textbf{Study 2}) that aims to evaluate the gestures produced by our model and its capacity to perform "style preservation". A third study (\textbf{Study 3}) was conducted to evaluate the style transfer of speakers seen during training - \emph{Seen Speaker} condition - , as well as speakers unseen during training - \emph{Unseen Speaker} condition. In these studies, we present a virtual agent simulated with the converted 3D poses of the 2D poses synthesized by our model. \textbf{Study 3} aims to assess the \emph{resemblance} of the produced stylized gestures to the target style. We additionally compare in \textbf{Study 3} our model's produced stylized gestures in \emph{Seen Speaker} condition, to Mix-StAGE that we consider our baseline.  
We used 7 factors linked to behavior expressivity to assess the quality of the 3D animation. We follow the recommendations proposed in  \cite{wolfert2021review} and assess on a 1 to 7 likert scale the first 5 factors: \emph{naturalness}, \emph{coherence}, \emph{human-likeness}, \emph{appropriateness}, and \emph{comprehensibility}. We add the 2 other factors \emph{synchronization}, and \emph{alignment} to evaluate the gestures' temporal property with speech.
In addition, 3 factors are used to evaluate the \emph{resemblance}, \emph{resemblance in terms of gestures amplitude}, and \emph{resemblance in terms of gestures dynamics} between the human gestures and the virtual agent's gestures.
We note that we distinguish between two types of factors that we want to assess in our studies: the first  ones (7 expressivity factors) are related to evaluating the virtual agent's \emph{behavioral expressivity} (for \textbf{Study 1} and \textbf{Study 2}), and the second ones (3 resemblance factors) are to assess the \emph{resemblance} of our model's stylized produced gestures with the ground truth (for \textbf{Study 1}) and with the target style (for \textbf{Study 3}). Each factor is rated on a 7 likert scale. 
30 participants are recruited for each study, including for the pre-evaluation study (\textbf{Study 1}), on Prolific, an online crowd-sourcing website. Participants are selected such that they are fluent in English and have a university degree. Attention checks are added in the beginning and the middle of each study to filter out inattentive participants. All the animations presented in these studies are produced on a 3D virtual agent.
\subsubsection{\textbf{Study 1: 3D Animation Pre-Evaluation}}
The first human perceptual study we conduct aims to assess our approach for the 2D to 3D pose conversion. In this study, we present 4 pairs of videos: for each pair, the first video shows the generated 3D poses simulated on a virtual agent, and the second one is the video of the original speaker performing the same gestures. The 2D poses that we use for this 3D conversion are ground truth data extracted from PATS. 
\subsubsection{\textbf{Study 2: Gesture Generation Evaluation}}
To assess the quality of the 2D poses generated by our model, and its ability to perform "style preservation" and remember the unique style of each speaker, we conduct another human perceptual study. We use the 7 expressivity factors that are used in the pre-evaluation study to assess the quality of the produced virtual agent's gestures.
This study consists of  8 videos: 4 videos show 3D animations of our model's predictions, and 4 other videos show the converted 2D to 3D poses animation of the original speaker's gestures which serve as ground truth. For each video, participants are asked to rate the 7 expressivity factors on a 1 to 7 likert scale \cite{wolfert2021review}.
\subsubsection{\textbf{Study 3: Style Transfer and Zero-Shot Style Transfer Evaluation}}
The third perceptive study aims to assess the style transfer correctness performed by our model for both conditions: \emph{Seen Speaker} and \emph{Unseen Speaker}. For each condition, participants watch 3 videos representing the ground truth (\emph{video 1}), the target speaker (\emph{video 2}) and our model (\emph{video 3}), respectively. We ask the participants to answer questions related to the 3 resemblance factors, provided in a random order.
For the \emph{Seen Speaker} condition, we present 12 videos: 4 videos show the 3D animation of the source speaker gestures, 4 other videos show the 3D animation of the target speaker gestures, and the remaining 4 videos show the simulation of our model's predictions in 3D, after performing the style transfer from the target speaker to the source speaker.
For the \emph{Unseen Speaker} condition, we present 9 videos (different videos from the above ones): 3 videos with the source speaker gestures, 3 with the target speaker gestures, and the remaining 3 with our model's 3D simulated predictions after performing the style transfer from target speakers not seen during training, to the source speakers. We note that in this experimental study, the \emph{resemblance} factors are the most important ones, since we want to assess the degree of resemblance of our model's stylized gestures to the target style.
For each set of questions in each condition, the target 3D animation is presented to the participants as a "baseline". We ask the participants to choose one of the two video - the source speaker 3D animation, and the 3D simulation of our model's predictions - that resembles the most to the baseline in terms of the 3 resemblance factors. Participants are asked the following questions: (1) Which video resembles the most to the baseline video ?; (2) Which video resembles the most to the baseline video in terms of gestures dynamics ?; and (3) Which video resembles the most to the baseline video in terms of gestures amplitude ?

\textbf{\emph{Comparing to the baseline \emph{Mix-StAGE}}: } We compare our stylized generated gestures in \emph{Seen Speaker} condition with the predictions of \emph{Mix-StAGE}\cite{ahuja2020style}, which serves as a baseline for this condition. We ask the participants to watch 3 videos representing the ground truth (\emph{video 1}), the target speaker (\emph{video 2}), and Mix-StAGE predictions after performing style transfer from target speakers to source speakers (\emph{video 3}). We repeat this question 3 times (presenting 9 videos in total), and assess the \emph{resemblance} of \emph{Mix-StAGE}'s produced gestures with respect to the target speakers.
\section{Results and Discussion}
\subsection{Objective Evaluation Results}
\begin{figure}[H]
\hspace{-1.cm}
\minipage{0.4\textwidth}
  \includegraphics[width=8cm, height=5cm,]{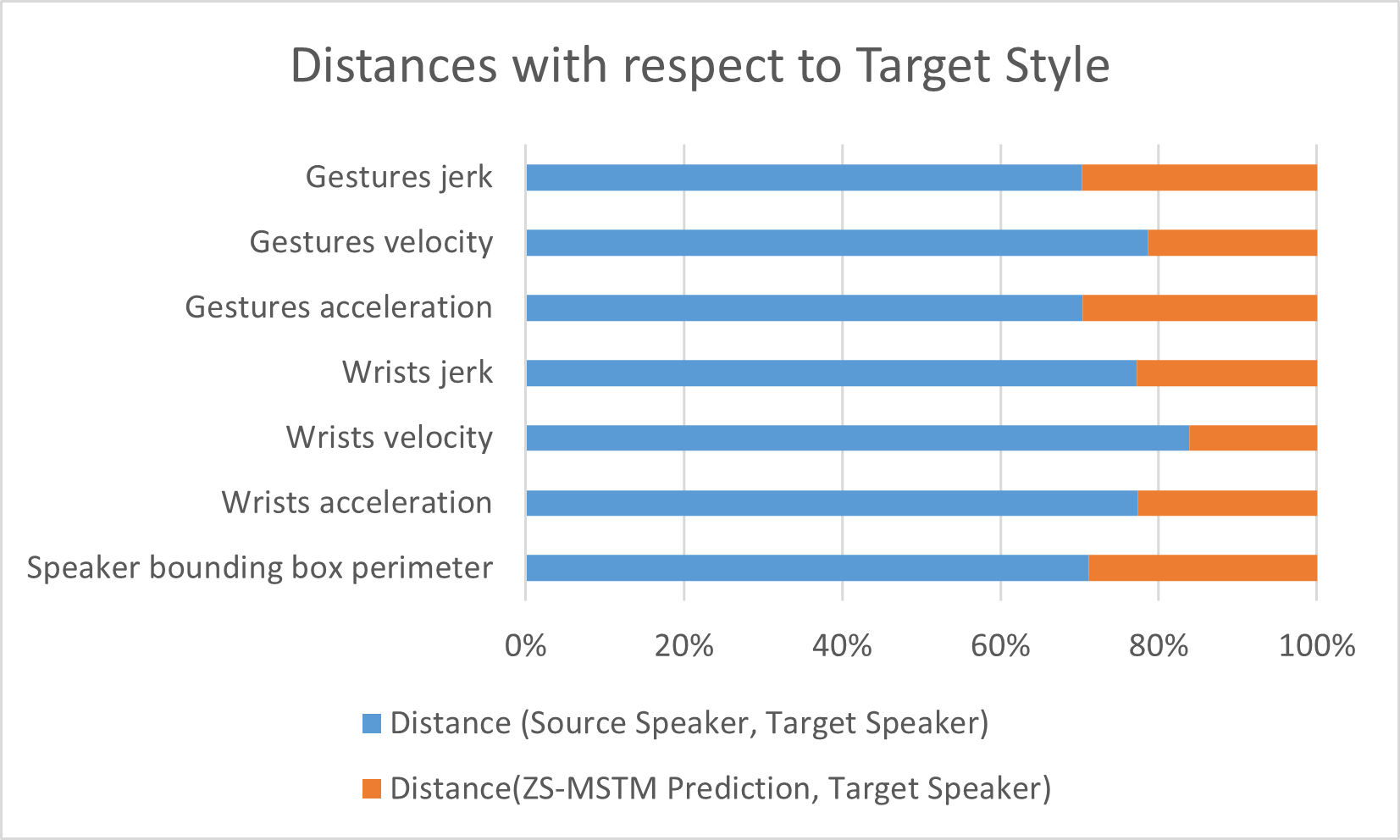}
\centerline{\hspace{2cm}  (a) style transfer from seen speakers}
  \label{fig:Distances_ST}
\endminipage\hfill
\minipage{0.4\textwidth}
  \hspace{-1.25cm}
  \includegraphics[width=8cm, height=5cm,]{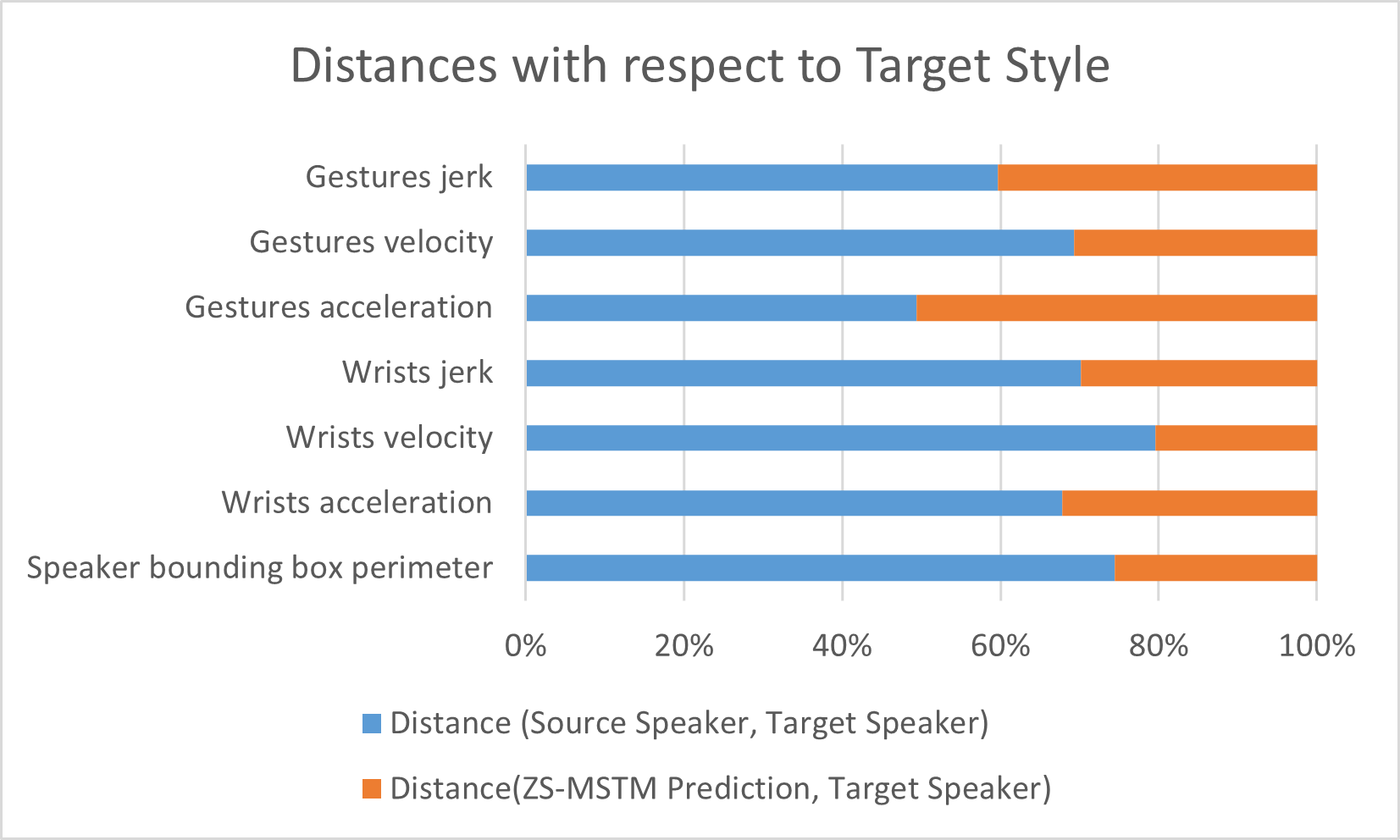}
   \centerline{(b) style transfer from unseen speakers}
  \label{fig:Distances_ZSST}
\endminipage\hfill
 \caption{Distances between the target speaker style and each of the source style and ZS-MSTM's generated gestures style for seen target speakers: on left, (a) style transfer using seen speaker during training; on right, (b) style transfer using unseen speaker during training.}
\label{fig:Distances}
\vspace{-0.3cm}
\end{figure}
Objective evaluation experiments are conducted for evaluating the performance of our model in the \emph{Seen Speaker} and \emph{Unseen Speaker} conditions. For \emph{Seen Speaker} condition, experiments are conducted on the test set that includes the 16 speakers that are seen by our model during training. For \emph{Unseen Speaker} condition, experiments are also conducted on another test set that includes the 6 speakers that were not seen during training.
For both conditions, we define two sets of distances: \textbf{\emph{Dist.(}}\emph{Source}, \emph{Target}\textbf{)} - representing the average distance between the source style and the target style on the corresponding test set -, and \textbf{\emph{Dist.(}}\emph{ZS-MSTM}, \emph{Target}\textbf{)} - representing the average distance between our model's gestures style and the target style -.  Fig. \ref{fig:Distances} (a) reports the experimental results on the \emph{Seen Speaker} test set. It illustrates the results of \textbf{\emph{Dist.(}}\emph{Source}, \emph{Target}\textbf{)} in terms of behaviors dynamics and speaker bounding box perimeter between the target speaker style and the source speaker style. Experimental results for the \emph{Unseen Speaker} test set are depicted in Fig. \ref{fig:Distances} (b). For \emph{Seen Speaker} condition (Fig. \ref{fig:Distances}(a)), \textbf{\emph{Dist.(}}\emph{Source}, \emph{Target}\textbf{)} is higher than 70\% of the total distance for all behavior dynamics metrics.; thus \textbf{\emph{Dist.(}}\emph{ZS-MSTM}, \emph{Target}\textbf{)}
is less than 30\% of the total distance for all behavior dynamics metrics. Wrists velocity, jerk and acceleration results reveal that the virtual agent's arms movements show the same expressivity dynamics as the target style (\textbf{\emph{Dist.(}}\emph{ZS-MSTM}, \emph{Target}\textbf{)} < 22\%).
The style transfer from target speaker "Shelly" to source speaker "Angelica" - knowing that Angelica is a \emph{Seen Speaker} - shows that the distance of predicted gestures' behavior dynamics metrics are close (distance < 20\%) to "Shelly" (\emph{target style}), while the ones between "Angelica" and "Shelly" are far (distance > 80\%). 
The perimeter of the prediction's bounding box (BB) is closer (distance < 30 \%) to the target speaker's BB perimeter than the source . The closeness between predictions dynamics behavior metrics values are shown for all speakers in the \emph{Seen Speaker} condition, specifically for the following style transfers - \emph{target} to \emph{source} - : "Fallon" to "Shelly", "Bee" to "Shelly", "Conan" to "Angelica", "Oliver" to "lec\_cosmic", which are considered having different lexical diversity, as well as spatial average extent, as reported by the authors of PATS \cite{ahuja2020style}.  

For the \emph{Unseen Speaker} condition, results reveal that our model is capable of reproducing the style of the 6 unseen speakers. As depicted in Fig. \ref{fig:Distances} (b), for all behavior dynamics metrics, as well as the bounding box perimeter, \textbf{\emph{Dist.(}}\emph{Source}, \emph{Target}\textbf{)} is higher than 50\% of the total distances for all metrics. Results show that for wrists velocity, jerk and acceleration, \textbf{\emph{Dist.(}}\emph{ZS-MSTM}, \emph{Target}\textbf{)} is less than 33\%. Thus, arm movement's expressivity produced by \emph{ZS-MSTM} is close to the one of the target speaker style. Moreover, the perimeter of the prediction's bounding box is close (distance < 30 \%) to the target speaker's, while the distance between the BB perimeter of the source and the target is far (distance > 70 \%). While our model has not seen "Lec\_evol"'s multimodal data during training, it is yet capable of transferring his behavior expressivity style to the source speaker "Oliver".  It is also capable of performing zero-shot style transfer from the target speaker "Minhaj" to the source speaker "Conan". In fact, results show that wrists acceleration and jerk values of our model's generated gestures are very close to those of the target speaker "Minhaj". We observe the same results for the 6 speakers for the \emph{Unseen Speaker} condition. 

We additionally conduct a Fisher's LSD Test to do pair-wise comparisons on all metrics, for the two set of distances - \textbf{\emph{Dist.(}}\emph{Source}, \emph{Target}\textbf{)}, and \textbf{\emph{Dist.(}}\emph{ZS-MSTM}, \emph{Target}\textbf{)} - in both conditions. We find significant results ($p< 0.003$) for all distances in both conditions. 
\subsection{Human Perceptual Studies Results}
\subsubsection{\textbf{Study 1: 3D Animation Pre-Evaluation}} 
The first human perceptual study is the pre-evaluation to assess the 3D data animation and simulation on a virtual agent, which are converted from the 2D generated poses. We calculate the mean values obtained on the 7 expressivity factors and on the 3 resemblance factors. 
Results show that all factors received a mean score above 3 on a likert-scale from 1 to 7. They reveal that the 2D to 3D conversion of the 2D-poses generated by our model tend to resemble the human's gestures which served as ground truth in this evaluation. We observe that the factor \emph{Resemblance} gets the highest mean (above 4) and that the factor \emph{Gestures Amplitude Resemblance} gets the highest second mean score, followed by the factor \emph{Naturalness}. This indicates that the 3D animations show gestures that resemble the human's gestures, especially in terms of gestural amplitude resemblance. We obtain similar mean scores (3.5$<$ \emph{mean} $<$3.6) for the factors \emph{Comprehensibility}, \emph{Gestures Dynamics Resemblance}, \emph{Likeness}, and \emph{Alignment}. The mean score for the remaining factors is 3.1. While the 3D pose animation has not received the highest possible rate, its results are nevertheless good enough to be used as ground truth. In the remaining evaluations, all animations are obtained with this method, offering similar behavior quality.
\subsubsection{\textbf{Study 2: Gesture Generation Evaluation}}
The second human perceptual study consists of assessing the quality of the generated poses and the ability of our model to perform "style preservation", thus its capacity of remembering the unique style of each speaker. We calculate the mean scores for the 7 behavioral expressivity factors. 
We observe that our model's predictions (\textbf{P}) get mean values that are close to those of the ground truth (\textbf{GT}), especially for the factors \emph{Appropriateness} (mean difference(\textbf{GT}, \textbf{P})=0.1) and \emph{Comprehensibility} (mean difference(\textbf{GT}, \textbf{P})=0.3). The remaining factors have higher mean difference between the ground truth and predictions: \emph{Coherence} (mean difference=0.4), \emph{Human-likeness} (mean difference=0.44),  \emph{Synchronization} (mean difference=0.5), \emph{Alignment} (mean difference=0.51), and \emph{Synchronization} (mean difference=0.53). We additionally perform a Fisher’s LSD Test to do pair-wise comparisons of the means of the 7 factors. Significant results ($p<0.001$) are found for the factors \emph{Appropriateness}, \emph{Comprehensibility}, \emph{Coherence} and \emph{Human-Likeness} when comparing values for the Ground Truth gestures those of our model's generated gestures. This constitutes experimental validation that our model is perceived significantly close to the ground truth, and therefore allows "style preservation". Therefore, our model is able to remember the unique style of each speaker, even though it is trained on multiple ones. While our model is perceived significantly close to the ground truth, results show that we still need to leverage the synchronization of the produced gestures with the speech and its content. 
\vspace{-0.2cm}
\subsubsection{\textbf{Study 3: Style Transfer and Zero-Shot Style Transfer Evaluation}}
\begin{table}
\centering
\small
\begin{tabular}{c|cc|cc|cc}
\hline
\textbf{Resemblance Metrics} &
  \multicolumn{2}{c|}{\textbf{ZS-MSTM - Seen Speaker}} &
  \multicolumn{2}{c|}{\textbf{ZS-MSTM - Unseen Speaker}} &
  \multicolumn{2}{c}{\textbf{Mix-StAGE}} \\ \hline \hline
\textbf{Resemblance to the target style} &
  \multicolumn{1}{c|}{\textit{\textbf{Source Style}}} &
  \textit{\textbf{Prediction}} &
  \multicolumn{1}{c|}{\textit{\textbf{Source Style}}} &
  \textit{\textbf{Prediction}} &
  \multicolumn{1}{c|}{\textit{\textbf{Source Style}}} &
  \textit{\textbf{Prediction}} \\ 
Globally &
  \multicolumn{1}{c|}{$0.35 \pm 0.02$} &
  $0.65 \pm 0.04$  &
  \multicolumn{1}{c|}{$0.46 \pm 0.01$} &
  $0.54 \pm 0.03$ &
  \multicolumn{1}{c|}{$0.57 \pm 0.03$} &
  $0.43 \pm 0.04$ \\ 
\begin{tabular}[c]{@{}c@{}} W.r.t. gesture dynamics\end{tabular} &
  \multicolumn{1}{c|}{$0.32 \pm 0.05$} &
  $0.68 \pm 0.05$ &
  \multicolumn{1}{c|}{$0.47 \pm 0.02$} &
  $0.53 \pm 0.05$ &
  \multicolumn{1}{c|}{$0.56 \pm 0.03$} &
  $0.44 \pm 0.04$ \\ 
\begin{tabular}[c]{@{}c@{}} W.r.t. gesture amplitude\end{tabular} &
  \multicolumn{1}{c|}{$0.42 \pm 0.03$} &
  $0.58 \pm 0.06$ &
  \multicolumn{1}{c|}{$0.42 \pm 0.04$} &
  $0.58 \pm 0.04$ &
  \multicolumn{1}{c|}{$0.54 \pm 0.05$} &
  $0.46 \pm 0.05$ \\ \hline
\end{tabular}
\vspace{0.25cm}
\caption{Results of the perceptual study for the conditions ZS-MSTM (seen speakers), ZS-MSTM (unseen speakers), and baseline (Mix-StAGE). We also report the confidence intervals.}
\label{Table_Results}
\vspace{-0.7cm}
\end{table}
The first four columns (\textbf{ZS-MSTM - Seen Speaker} and \textbf{ZS-MSTM - Unseen Speaker}) of Table \ref{Table_Results} shows the results of the human perceptual study for assessing the stylized gestures generated by our model for both conditions \emph{Seen Speaker} and \emph{Unseen Speaker}. Results show that, on a scale from 0 to 1 representing the number of times our model is selected to resemble the target style, our model's predictions get values above 0.58 for condition \emph{Seen Speaker}, and values between 0.53 and 0.58 for condition \emph{Unseen Speaker}. 
Our model's generated style in condition \emph{Unseen Speaker} is perceived as having quite high resemblance to the target style (score of 0.54), especially in terms of gesture amplitude (score of 0.53) and gesture dynamics (score of 0.58). We additionally performed t-test comparison between source style values and prediction style scores for the conditions \emph{Unseen Speaker} and \emph{Seen Speaker}. Significant results ($p<0.001$) are found  between the Source scores and the Prediction scores. These results reveal that our model's generated stylized gestures are significantly perceived as being closer to the target style than to the source style.  

\textbf{\emph{Comparing to the baseline \emph{Mix-StAGE}}: } The first two columns (\textbf{ZS-MSTM - Seen Speaker}) and the last two columns (\textbf{Mix-StAGE}) of Table \ref{Table_Results} present the results when comparing our generated gestures in condition \emph{Seen Speaker} with the baseline \emph{Mix-StAGE} which only operates in this condition and not in the condition \emph{Unseen Speaker}. On a scale from 0 to 1 representing the number of times our model is selected to resemble the target style, our model gets scores between 0.58 and 0.65, while Mix-StAGE gets lower scores, between 0.43 and 0.46. We additionally conduct a Fisher LSD test to do pair-wise comparisons of the means between the 3 factors of both conditions \emph{Mix-StAGE} and \emph{ZT-MSTM}, and identify the cases where the means are statistically different. We find a significant difference (p $<$ 0.003) for the factor \emph{Resemblance in terms of gesture dynamics}, and \emph{Resemblance in terms of gesture amplitude.}
\vspace{-0.3cm}
\section{Conclusion and Future Work}
We have presented the first approach for zero-shot multimodal style transfer for 2D pose synthesis that allows the transfer of style from any speakers unseen during the training phase. To visualize the resulting animation of a virtual agent, we have developed a model that converts 2D poses into 3D poses. Adding human mesh on the 3D poses allows us to simulate the 3D behavior of a virtual agent. Objective and subjective evaluations show that our model produces stylized animations that are close to the target speakers style even for unseen speakers. We have evaluated our model using behavior expressivity metrics as well as perceptive factors. In a next future, we plan to expand our model to consider dialog acts, and other semantic information to model more specifically gesture types (deictic, iconic, and metaphoric). In addition, we want to extend our style model to cover facial expressions and head movements.

\bibliographystyle{ACM-Reference-Format}
\bibliography{bib}
\appendix
\end{document}